\documentclass[fleqn,usenatbib]{mnras}

\usepackage[normalem]{ulem}
\usepackage{orcidlink}

\usepackage[T1]{fontenc}
\DeclareRobustCommand{\VAN}[3]{#2}
\usepackage{xcolor}

\let\VANthebibliography\thebibliography
\def\thebibliography{\DeclareRobustCommand{\VAN}[3]{##3}\VANthebibliography}

\usepackage{graphicx}	
\usepackage{amsmath}	
\usepackage{amssymb}	
\usepackage{newtxtext,newtxmath}

\title[]{Detection of two additional circumbinary planets around Kepler-451}

\author[Ekrem M. Esmer et al.]{
Ekrem Murat Esmer$^{1}$\thanks{E-mail: esmer@ankara.edu.tr}\orcidlink{0000-0002-6191-459X},
Özgür Baştürk$^{1}$\orcidlink{0000-0002-4746-0181},
Selim Osman Selam$^{1}$\orcidlink{0000-0002-4953-4818},
and Sinan Aliş$^{2,3}$\orcidlink{0000-0002-6990-8899}
\\
$^{1}$Department of Astronomy and Space Sciences, Faculty of Science, Ankara University, 06100 Ankara, Turkey\\
$^{2}$Department of Astronomy and Space Sciences, Faculty of Science, Istanbul University, 34119 Istanbul, Turkey\\
$^{3}$Istanbul University Observatory Research and Application Center, 34119 Istanbul, Turkey
}

\date{Accepted 2022 February 4. Received 2022 February 3; in original form 2022 January 20}

\pubyear{2021}

\begin{document}
\label{firstpage}
\pagerange{\pageref{firstpage}--\pageref{lastpage}}
\maketitle

\begin{abstract}
We announce the detection of two new planetary-mass companions around Kepler-451 binary system in addition to the one detected previously based on eclipse timing variation analysis. We found that an inner planet with 43 d period with a minimum mass of 1.76 M$_{\rm jup}$ and an outer one with a $\sim$1800 d orbital period with a minimum mass of 1.61 M$_{\rm jup}$ can explain the periodic variations in the residuals of the one-planet fit of the eclipse timings. We updated the orbital period of the middle planet as 406 d, and determined its eccentricity as 0.33. The newly discovered outer planet is also on an eccentric orbit (0.29), while the innermost planet was assumed to have a circular orbit. All three Jovian planets have similar masses, and our dynamical stability test yields that the system is stable.

\end{abstract}




\section{Introduction}

Eclipsing post-common envelope binary systems with a subdwarf B-type primary (sdB) and an M-dwarf secondary (dM) offer a favourable observational opportunity to determine eclipse timings within very good precision. Besides the short-lived eclipse events owing to the small sizes of its components, the primary eclipse, in particular, is also deep, symmetrical and V-shaped due to the large temperature contrast making the measurements of the profile center even more efficient. Therefore, the variations in their eclipse timings, also known as ETVs, can be studied in more detail based on more precise and accurate datasets of minima timings. Moreover, additional bodies with substellar masses might have formed in an accretion disk built up by the material lost from the upper layers of the progenitor of the sdB component \citep{zorotovic2013} in the short duration following the ejection of the common envelope. In fact, such potentially "second-generation" planets have been suggested in multiple sdB + dM binaries (HW Vir: \cite{lee2009}, NN Ser: \cite{beuermann2010}, NY Vir: \cite{lee2014}). Nevertheless, \cite{bearsoker2014} came up with an alternative explanation for the survival of the first-generation circumbinary (CB) planets from the ejection of the common envelope too. An orbital decay can be expected from such systems due to a resonant interaction between the binary and a CB disk \citep{chenpod2017}, the evidence of which was observed in the NN Ser system \citep{hardy2016}.

Kepler-451 (KIC 9472174) is an sdB + dM eclipsing binary displaying a strong reflection effect on the secondary caused by the very high surface temperature of the primary on its light curves. \cite{ostensen2010} was first to analyze the Kepler light curve of the system as well as its radial velocities. They reported the temperature of the primary as 29564 K from their spectroscopic analyses. They also derived the mass of the sdB component by making use of the surface gravity value (log g) from spectroscopy in a mass-radius calibration as 0.48 M$_\odot$, which is the canonical mass for an sdB pulsator. Based on this value, they were able to compute the secondary mass as 0.12 M$_\odot$ from the light curve analysis.

The first planet detected in the Kepler-451 system was announced by \cite{Baran2015} based on its ETVs. The suggested Jovian-mass planet was reported to have an orbital period of 416 days about the common center of mass with the binary pair. In a following study, \cite{Krzesinski2020} were not able to detect the signal and attributed that found by \cite{Baran2015} to a consequence of the data extraction and processing. Therefore, the planet's existence is disputed. Subsequently, a research report from \cite{Baran2020RNAAS} claimed the discovery to be valid by recalculating the timings and showing that the ETV signal is indeed periodic.

Kepler space telescope observed the system in its primary mission continuously for almost four years. This dataset forms the backbone of all the studies of the system. \cite{Baran2015} also reported ETVs either due to a secular or a long term cyclic change that could not be covered by the time-span of the Kepler data. The variation in that time range resembles an upward parabola, which is not expected considering the lack of mechanisms that would lead to either mass-loss from the secondary or mass-accretion by the primary. On the other hand, another potentially bound, yet unseen CB object can explain the observed part of a variation with a period longer than the time-span of the observations. Nevertheless, interactions between the objects and a potential CB disk could be more complicated than theorized.

With the arrival of TESS observations of the system, we investigated this question that long waited to be answered. We modelled the light curves from Kepler and TESS, as well as from our own observations to extend the wavelength coverage and to track the ETVs in the large data gaps between TESS and Kepler observations. We used these models to calculate the mid-eclipse times, which we also derived from the well-established Kwee-van Woerden technique \citep{kwee1956} for comparison, and performed a detailed ETV analysis of the system. We were able to reveal two additional periodic signals in ETVs, which we attribute to two newly found planetary-mass bodies orbiting the binary pair. These discoveries make Kepler-451, the first system with three CB planets in an sdB + dM system, and second to Kepler-47 \citep{orosz2012} when all CBs are considered. Finally, we carried out a dynamical stability test for this interesting system, which can open up a new window in the discussion on their formation, whether the numbers, distances, and total masses of the detected bodies can be explained by the ejection of the common envelope material or not, assuming the observed ETVs are due to the suggested unseen second-generation additional bodies. We also investigate if potential magnetic-activity induced orbital period variations can be an alternative or complimentary explanation \citep{applegate1992}.

\section{Observations and Data}\label{sec:ObsData}

\begin{figure}
    \includegraphics[width=\columnwidth]{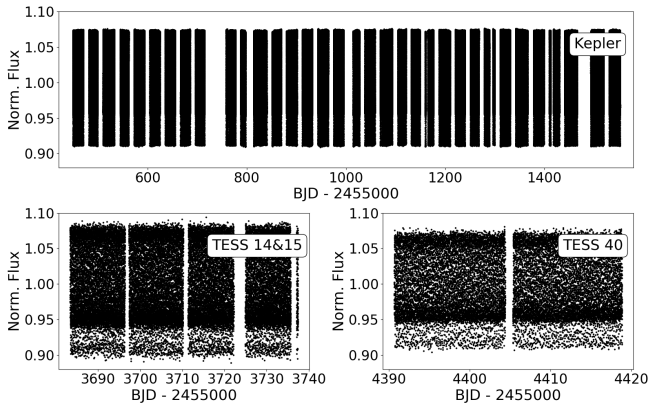}
    \caption{Detrended and normalized light curves of Kepler-451 that were used for producing the phase-folded data. Top panel shows the light curves from Kepler. The bottom left panel shows TESS light curves from sector 14 \& 15, and the bottom right panel shows sector 40. Note that only Kepler and TESS sector 14 \& 15 were used for the light curve analysis. See text for more details.}
    \label{fig:lcs}
\end{figure}

We gathered Kepler and TESS data through Space Telescope Science Institute's (STScI)  Mikulski Archive for Space Telescopes (MAST)\footnote{https://archive.stsci.edu} portal service. The Kepler dataset has a time range from 2009 to 2013. We used only the short cadence pre-search data conditioned simple aperture photometry (PDC-SAP) data that are more suitable for the light curve analysis compared to the long cadences in which the eclipses are not even detectable due to the short orbital period. We removed the trends on each observational quarters by applying third degree polynomial fits. Since both ends of almost all the quarters have strong linearity issues, we had to discard corresponding data points as well as random outliers before the detrending procedure. We normalized each cycle to their individual median value (see Fig. \ref{fig:lcs}). For the light curve analysis, we phase-folded these data and binned them to have equally-spaced 300 points for an orbital cycle.

TESS observed the system in sectors 14, 15 and 40. We applied the same detrending, phase-folding and normalization procedures that we did for the Kepler data, but this time on each sector separately. The trends in the two parts of each sector separated by a gap due to data downlink seem to be quasi-independent. Therefore, we divided the sectors into two parts from the data gap, where the telescope did not observe its targets. After the normalization, we initially realized that the amplitude of the eclipse light curve is similar in the first two sectors, whereas it changes in the sector 40 (see Fig. \ref{fig:lcs}). Therefore, while we used all TESS data to derive mid-eclipse timings, we only made use of the light curves from sectors 14 and 15 for the light curve modeling in Section \ref{sec:lc_analysis} because they form a continuous light curve with more data points and excluded that acquired during the sector-40 from that analysis only.

To provide a wider wavelength coverage for the light curve analysis, as well as additional ground-based eclipse timings especially in the large gaps between space-borne observations, we observed the binary with i) brand new 80-cm telescope (T80) at Ankara University Kreiken Observatory (AUKR), ii) with the 1-m telescope (T100) in TUBITAK National Observatory (TUG) and iii) with Istanbul University's 60-cm telescope (IUO) at \c{C}anakkale Onsekiz Mart University Ulup{\i}nar Observatory. These observations were dark, bias and flat-field corrected and the differential photometry has been performed by using {\sc AstroImageJ} software \citep{AstroImageJ}. The log of the observations and some related statistics are provided in Table \ref{tab:log_obs}.
We selected the observations from AUKR on the dates of 16, 17, 20 and 21 Aug 2021 for the light curve modeling due to their smaller scatter and full orbital phase coverage. The remaining observations were used only for the calculation of eclipse timings due to inferior photometric quality caused by unfortunate weather conditions or insufficient phase coverage.

\begin{table}
    \centering
    \caption{Log of photometrical data of Kepler-451. The observations marked with asterisks are used in the light curve analysis. For Kepler and TESS, observation dates and photometric scatter ($\sigma_{\rm phot}$) are for the phase folded and binned data, while the exposure times are the time between two binned data points.}
    \begin{tabular}{c|c|c|c|c}
       Date of Obs.&Filter &Exp. (s)& $\sigma_{\rm phot}$ (mmag) & Telescope \\ \hline
        12 Jun 2017& R & 12 & 4.6 & TUG\\
        10 May 2018& R & 25 & 3.2 & TUG\\
        19 Jul 2018&R&50&3.5&TUG\\
        31 May 2019&R&25&5.9&TUG\\
        09 Jun 2019&R&45&3.5&TUG\\
        10 Nov 2019&R&25&4.7&TUG\\
        &V&25&4.2&TUG\\
        19 Jul 2020&R&70&5.8&IUO\\
        07 Oct 2020&R&25&5.0& TUG\\
        23 Apr 2021&R&24&4.6&TUG\\
        16 Jul 2021&R&30&5.9&IUO\\
        19 Jul 2021&B&45& 3.3 & TUG\\
        21 Jul 2021&V&80& 3.8 & TUG\\
        29 Jul 2021&R&30&6.0&IUO\\
        05 Aug 2021&V&80&4.0&TUG\\
        16 Aug 2021*&$g'$&40& 1.8 & AUKR\\
        17 Aug 2021*&$r'$&55& 2.5 & AUKR\\
        20 Aug 2021*&$i'$&80& 3.0 & AUKR\\
        21 Aug 2021*&$g'$&40& 2.0 & AUKR\\
        22 Aug 2021&$r'$&50& 3.8 & AUKR\\
        08 Sep 2021&$g'$&40&3.4& TUG\\
        18 Sep 2021&$g'$&45&3.9& TUG\\
        21 Sep 2021&R&50&4.8& TUG\\
        27 Sep 2021&Clear&30&4.1& TUG\\ \hline
        15 Oct 2011* & - & 32.5 & 0.17 & Kepler \\
        13 Aug 2019* & TESS & 32.5 & 0.75 & TESS\\
    \hline
    \end{tabular}
    \label{tab:log_obs}
\end{table}

\cite{ostensen2010} published radial velocity observations of the system. Their spectra is single lined, therefore the radial velocity data are only for the primary component. We gathered this dataset from CDS's Vizier archive service\footnote{https://cdsarc.cds.unistra.fr/} and used it during the analysis, which we describe in the next section.

\section{Light Curve Modeling}\label{sec:lc_analysis}

The light curves of the binary display a strong reflection effect around the secondary eclipse, due to the large surface temperature contrast and the small orbital separation between the components. While the primary eclipse is deeper than the secondary, the depths of both eclipses are smaller compared to the amplitude of the reflection, most probably due to relatively low orbital inclination. Although it is not easy to notice with the eye, the sdB component's pulsations are detectable in Kepler light curves and have been reported by \cite{ostensen2010}. Since we binned the Kepler and TESS light curves to 300 points each, we i) reduced the computation time, and ii) removed the effects of the pulsations of the sdB star. Therefore, the assumption that the light curves represent only the eclipses and phase-dependent modulations is safe.

Because of the radial velocity of the secondary is missing and the eclipses are not total, the mass ratio and related parameters are degenerate. Therefore we adopted the mass value from \cite{ostensen2010}, as 0.48 M$_\odot$, which they derived by using both the spectral $logg$ and the orbital solution, in the range for post common envelope sdBs. Since, we adopted the mass of the primary, the mass ratio of the system can be calculated easily from the radial velocities as $q$ = 0.25, and the mass of the secondary is 0.12 M$_\odot$. Semi-major axis of the binary was also calculated as 0.891 R$_\odot$.

We adopted and fixed the primary's temperature to $29564\ K$ that \cite{ostensen2010} reported, which is comparable with the value calculated by \cite{lei2018_teff1} from fitting the hydrogen and the helium lines with synthetic spectra. Other than the parameters mentioned above, we fixed the gravity brightening values (g$_1 = 1$, g$_2 = 0.32$) and the albedo of the primary (A$_1 = 1$) to their theoretical values. We set all of the remaining parameters to freely vary. These parameters are, the temperature of the secondary (T$_{\rm eff,2}$), radii of the components (R$_{1}$ and R$_{2}$), albedo of the secondary (A$_2$), linear limb darkening coefficients, orbital inclination ($i$), and the passband luminosities. The binary has a very small separation and a short orbital period. Therefore, we assumed synchronous rotation for both of the components.

\begin{figure*}
    \includegraphics[width=460pt]{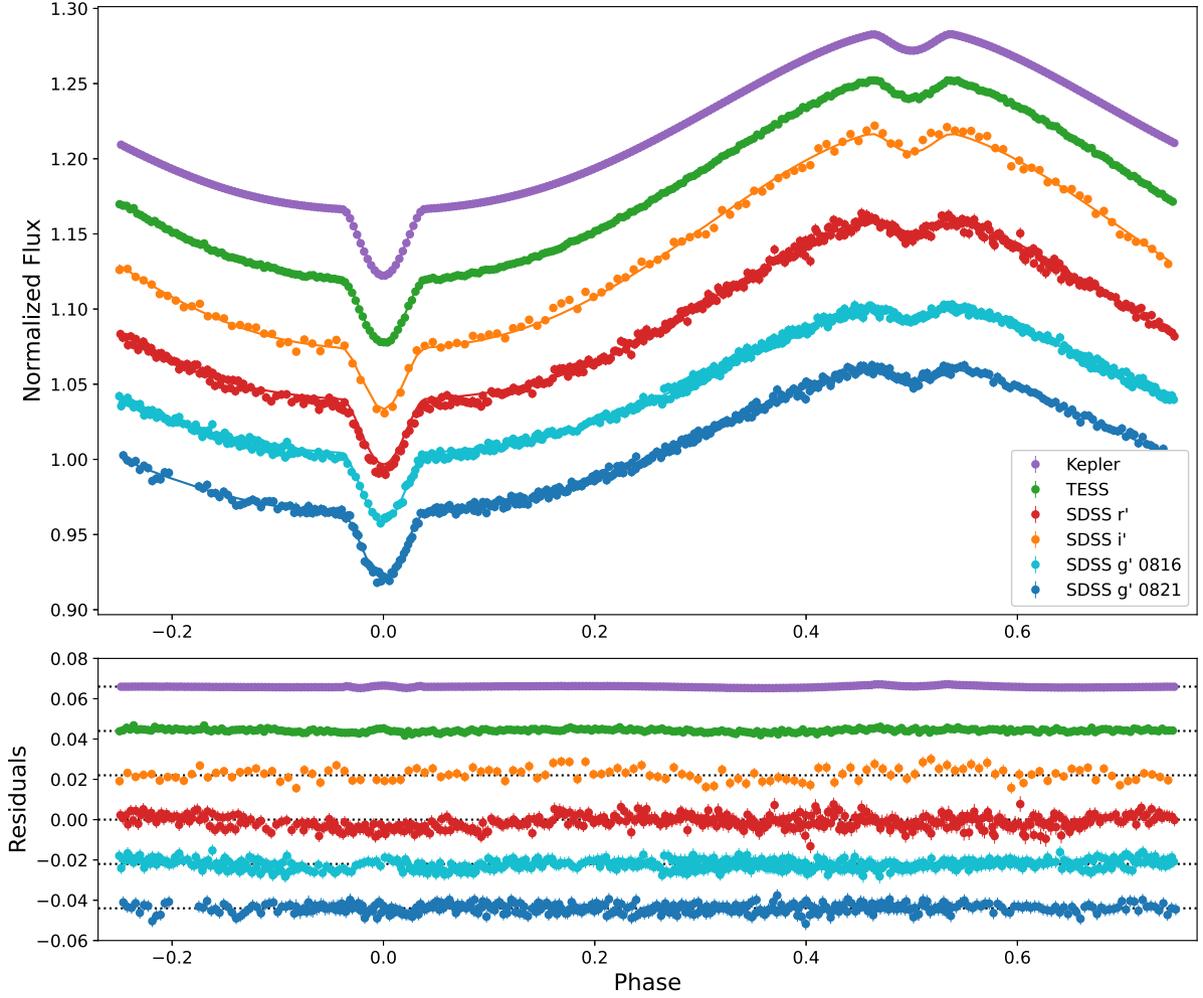}
    \caption{Light curves of Kepler-451 along with synthetic models from {\sc Phoebe}v2 and their residuals. Light curves and the residuals are arbitrarily shifted and color assigned.}
    \label{fig:lcmodel}
\end{figure*}

We used the second version of {\sc Phoebe} \citep{phoebe} for the modeling of the light and radial velocity curves. We first used downhill simplex algorithm \citep{neldermead_downhill} to find an initial global solution that resembles the observations. We then used Levenberg-Marquardt method (LM) \citep{Levenberg.1944, marquardt:1963} for fine-tuning the parameter values. However, we failed to calculate uncertainties of the parameters, most probably due to the small number of iterations that otherwise would have taken considerably longer computation times. To calculate the uncertainties, as well as to confirm the results of the light curve modeling, we also used the first version of {\sc Phoebe} \citep{phoebeLegacy}, by keeping the fixed and free parameters as they were in the modeling with the second version.

The best-fit parameter A$_2$ is considerably different for the analyses with the two different versions of {\sc Phoebe}. To investigate this difference, we made several fitting trials for different values of A$_2$ with the first version of {\sc Phoebe}. When we fixed A$_2$ to unity, the value of T$_{\rm eff,2}$ becomes 2987 K. On the other hand, the best-fit value of A$_2$ is 0.1520(8) and T$_{\rm eff,2}$ is 2829(1) K. Therefore the difference of the best-fit values of A$_2$ in the modeling attempts with different versions of {\sc Phoebe} turned out to be due to high parameter degeneracy between T$_{\rm eff,2}$ and A$_2$.

Except A$_2$, all the other best-fit parameter values from the two versions of {\sc Phoebe} are almost the same (see Table \ref{tab:lc_results}). Our results have relatively small discrepancies with the observations that can be seen in Fig. \ref{fig:lcmodel}, on the entire orbital cycle and also around the eclipses. Since our aim is to calculate the eclipse timings from the light curve model, but we could not improve them any better, we scaled the eclipse model to the observations that will be explained in the next section.

\begin{table}
    \centering
    \caption{Results from the light curve analysis of Kepler-451. The values marked with asterisks are fixed.}
    \begin{tabular}{c|c|c|c}
        Parameter &\textbf{{\sc Phoebe2}} & \textbf{{\sc Phoebe Legacy}}\\ \hline
        $i\ (^\circ)$ &70.1 & 69.92(3)\\
        T$_1\ (K)$&29564* & 29564*\\
        T$_2\ (K)$&2856&2829(1)\\
        $q$&0.25*&0.25*\\
        $a\ (R_\odot)$&0.891*&0.891*\\
        $M_1\ (M_\odot)$&0.48*&0.48*\\
        $M_2\ (M_\odot)$&0.12*&0.12*\\
        $R_1\ (R_\odot)$&0.205&0.203(1)\\
        $R_2\ (R_\odot)$&0.164&0.168(1)\\
        $log~g_1\ (cgs)$&5.496&5.508(12)\\
        $log~g_2\ (cgs)$&5.086&5.067(27)\\
        $g_1$&1*&1*\\
        $g_2$&0.32*&0.32*\\
        A$_1$&1*&1*\\
        A$_2$ &0.9566&0.1520(8)\\
        $L_1 / L_T (Kepler)$&0.999790&0.99889(4)\\
        $L_1 / L_T (TESS)$&0.999286&0.99808(37)\\
        $L_1 / L_T (g')$&0.999971&0.9999(16)\\
        $L_1 / L_T (r')$&0.999786&0.9992(19)\\
        $L_1 / L_T (i')$&0.999190&0.9975(28)\\
        $L_2 / L_T (Kepler)$&0.000210&0.00111\\
        $L_2 / L_T (TESS)$&0.000714&0.00192\\
        $L_2 / L_T (g')$&0.000029&0.0001\\
        $L_2 / L_T (r')$&0.000214&0.0008\\
        $L_2 / L_T (i')$&0.000810&0.0025\\
    \hline
    \end{tabular}
    \label{tab:lc_results}
\end{table}

\section{Eclipse Timings Analysis}\label{sec:timing}

In order to determine the mid-eclipse times, we fitted our light curve model cycle by cycle to each light curve by varying only the conjunction times while keeping all the other fit parameters unchanged. To compensate the small changes on the normalization level of each cycle, we added another parameter that shifts the brightness level of the model. We calculated mid-eclipse timings of our own observations, but we did not include them in the final ETV analysis since we found them not to be sufficiently precise for a meticulous analysis at the end. Therefore, our dataset for the final ETV analysis consisted of only Kepler and TESS observations thanks to their superior photometric precision. The secondary eclipses are considerably shallow, which makes timing measurements imprecise, leading to a larger scatter on the ETV diagram. Therefore, we only used primary minima for the ETV analysis.

\begin{figure}
    \includegraphics[width=\columnwidth]{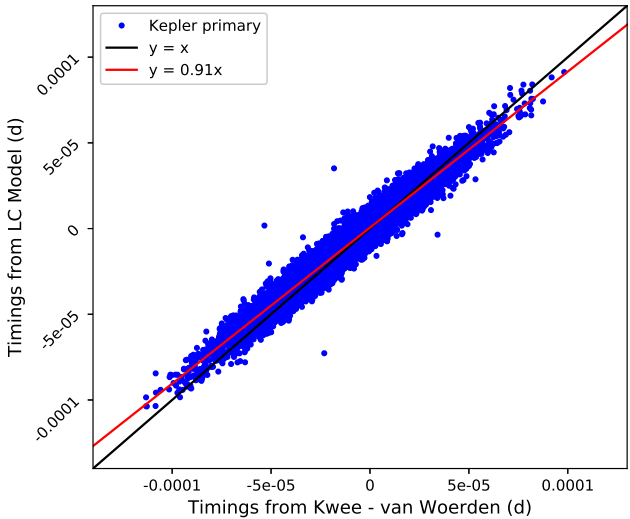}
    \caption{Comparison of primary mid-eclipse timings of Kepler data from widely used Kwee-van Woerden method (x-axis) and light curve model fitting method (y-axis). The red line shows the best-fit line to data, which has a slope of 0.91, whereas the black line shows the one-to-one relation.}
    \label{fig:kw_model_comp}
\end{figure}

We also measured mid-eclipse timings from widely used Kwee-van Woerden method (KW) \citep{kwee1956} for a comparison with the model-dependent mid-eclipse timings, which provided similar results. However, timings we derived from light curve models scattered within a narrower range than that calculated with KW, which is valid for both Kepler and TESS data, as well as for primary and secondary minima. We plotted the measurement results of primary eclipses from Kepler with respect to each other (Fig. \ref{fig:kw_model_comp}) for an illustration. As a result, we decided to base our timing analysis on the times of minima derived from light curve model fitting.

For the Kepler subset of our data, the overall trend is similar to what \cite{Baran2015} showed, hinting a cyclic variation on the order of hundred days and a longer-term variation, that resembles a secular increase on the period. TESS subset, on the other hand, has more evidence for the long-term variation to be cyclic. The trend is a decreasing one during the sectors 14 and 15, but turns and changes to an increase during the sector 40 supported by the positions of the most recent timings from our own observations. This smooth change indicates that the long-term variation in the Kepler data has a periodic nature.

We then performed a frequency analysis to only primary mid-eclipse timings in the Kepler dataset. Other than the orbital period frequency and its harmonics, we initially encountered a strong peak at $\sim$ 3.7 d and with an amplitude of $\sim$ 2.8 s in the Lomb-Scargle periodogram \citep{lomb1976, scargle1982} of the ETV diagram. This variation can easily be noticed with the eye and produces a significant fraction of the scatter in the Kepler data. Assuming the reason for this variation is the light-time effect (LiTE), its source will be unseen third body with a minimum mass was of  $\sim$ 0.044 M$_\odot$ on an orbit with a minimum semi-major axis of $a$ = 0.038 au. When we phase the ETV diagram with this period, data points clump at regular intervals, indicative of a beat phenomenon between sampling and periodicity. We found out that this clumping emerges in the secondary Kepler minima as well with a phase shift of 0.5, which is not observed in the TESS observations at all. This signal with 3.7 d period is in fact a spurious one caused by the sampling of Kepler data and the pulsations of the sdB component \citep{barlow2012ApJ...753..101B}. Since it has no real physical origin, we ignore this signal totally in our attempt to model the mid-eclipse timing data.

\subsection{Two long period circumbinary planets}\label{sec:frequency}

We first assumed two LiTEs to model i) the periodic variation that \cite{Baran2015} interpreted as a circumbinary planet and ii) the longer-term variation. The magnetic stellar winds are expected to remove mass from the sdB primary, hence decrease the orbital period of the binary. Therefore, we also assumed a model with a secular change superimposed on two LiTEs (see Section 4.2 in \citealt{esmer_hwvir} for formulas and definitions of ETV models). We used our own Python code for ETV modeling. We used Markov Chain Monte Carlo (MCMC) method for finding the global solution, by making use of the \emph{emcee} code \cite{emcee_foremanmackey2013}. We used Gaussian likelihoods and priors and set 100 walkers and 2000 random walks in total. After finding the global minimum of our model by taking the mode of the distributions of each parameter, we set them as initial values for fine-tuning their values with down-hill simplex and LM algorithms.

\begin{figure*}
    \includegraphics[width=460pt]{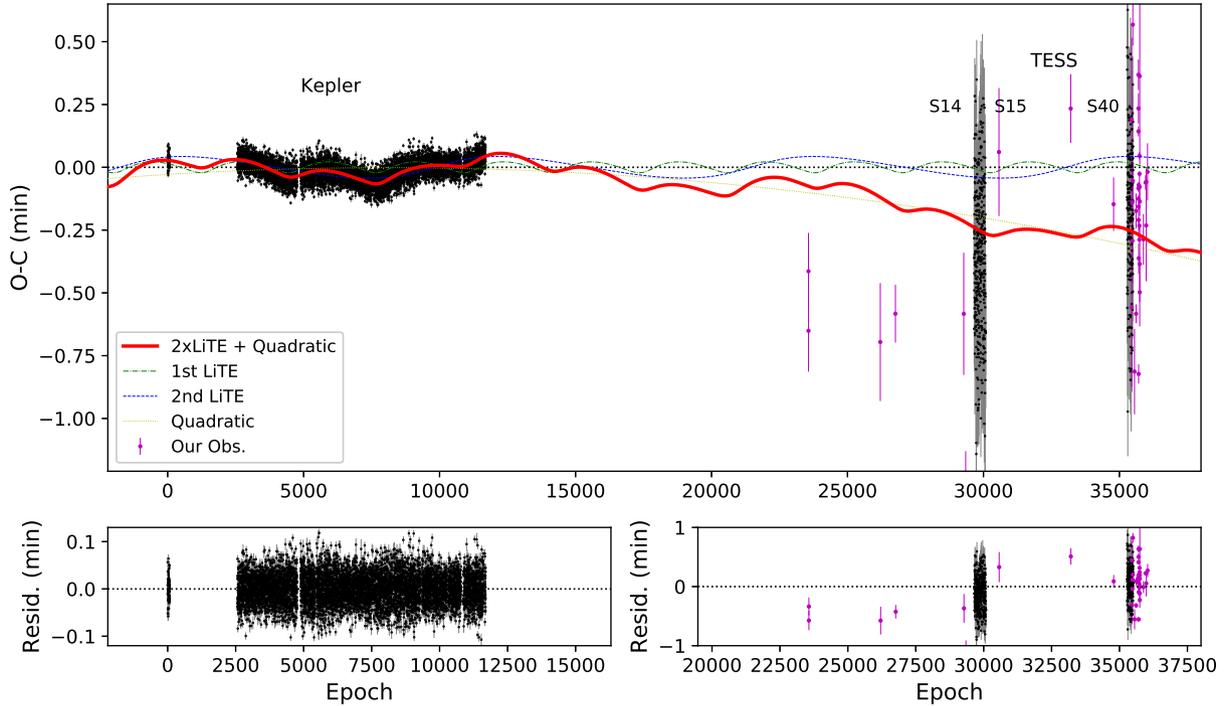}
    \caption{Mid-eclipse timings of Kepler-451 along with quadratic and two planet model (solid red). Each model component shown with different colors and line styles. Our observations from TUG, AUKR and IUO are shown in magenta. Note that the residuals divided in two regions and shown with different scales.}
    \label{fig:oc_model}
\end{figure*}

The best-fit solution for the first LiTE component of our model is similar to what \cite{Baran2015} found (see Table \ref{tab:etv_results}). We also found that this circumbinary planet with a minimum mass of 1.86 M$_{\rm jup}$ should have an eccentric orbit ($e$ = 0.33) although the period value we found (406 d) is slightly smaller than theirs (416 d). The longer period ($P$ = 1460 d) LiTE, on the other hand, corresponds to another circumbinary planet with a similar minimum mass (1.61 M$_{\rm jup}$), on a wide ( a$\ \sin i$ = 2.1 au) and eccentric orbit ($e$ = 0.29) (see Fig. \ref{fig:oc_model}). The uncertainty on the long-term LiTE is large, likely due to insufficient data coverage in one cycle, and heteroscedasticity of the dataset, which is formed by a combination of Kepler and TESS observations.

The root mean square (rms) of the residuals of the two planet solution is 4.85 s. The 3.7 d periodic signal is ignored as explained above, however, it affects the goodness of fit statistics of our best-fit solution. The $\chi_\nu^2$ of the best-fit is 12.625 for a degree of freedom of 8970, which is usually interpreted as either a poor fit or a result of underestimated observational errors. If we remove 3.7 d signal by fitting a \emph{sine} function, the $\chi_\nu^2$ decreases to 9.332, which is still not statistically satisfactory. This is, again, likely due to either poor representation of the \emph{sine} function or underestimated timing uncertainties. We leave this discussion here and adopt our best-fit two planet solution.

\subsection{Other potential explanations of the ETV}\label{sec:otherexplanations}

To explain whether the ETVs of Kepler-451 is caused by the magnetic activity of the cooler component we used the online tool\footnote{http://theory-starformation-group.cl/applegate/index.php} from \cite{voelschow2016A&A...587A..34V} that calculates the energy required to produce the quadrupole moment based on three different models, i) finite-shell two zone model, ii) finite-shell constant density model \citep{voelschow2016A&A...587A..34V}, and iii) thin shell model \citep{tian2009Ap&SS.319..119T}. For all three models, the ratio of the required energy ($\Delta E$) to the available energy for the magnetically active star ($E_{\star}$) has to satisfy the condition $\Delta E / E_{\star} << 1$. By using the value of the coefficients $k_1$ and $k_2$ (see the formulation in \citealt{voelschow2016A&A...587A..34V}) as 0.133 and 3.42, we calculated the energy required for the cooler component to derive \cite{applegate1992} mechanism based on the three models mentioned above. Considering the 1460 d periodic modulation, we calculated the minimum value of $\Delta E / E_{\star}$ for finite-shell two zone model as $\sim 12$, for finite-shell constant density model as $10^3$ and for the thin shell model as $\sim 0.7$. We also calculated the same parameter for the case of 406 d modulation as 143 for finite-shell two zone model, $10^3$ for finite-shell constant density model and 8 for the thin shell model. Since the magnetic activity of the dM component should have weakened significantly over the time it takes for the progenitor of the sdB component to have evolved from the main-sequence, Applegate mechanism is not expected to be effective in causing quadruple moment changes that lead to ETVs. In addition, magnetic-activity induced ETVs have more of a cyclic nature rather than a periodic one, with changing amplitudes from one cycle to another. Color, brightness, temperature, and maxima-level differences expected from strong magnetic activity variations are not observed in the light curves of the system either. Therefore, we concluded that the cyclic ETVs of Kepler-451 can not solely be explained by magnetic activity modulation. On the other hand, the orbit of the binary is circular, and the secondary minima are not out-of-phase with the primaries, hence apsidal motion can not be a possible explanation to the ETV of the system.

\subsection{The third planet on a smaller orbit}\label{sec:thirdplanet}

\begin{figure}
    \includegraphics[width=\columnwidth]{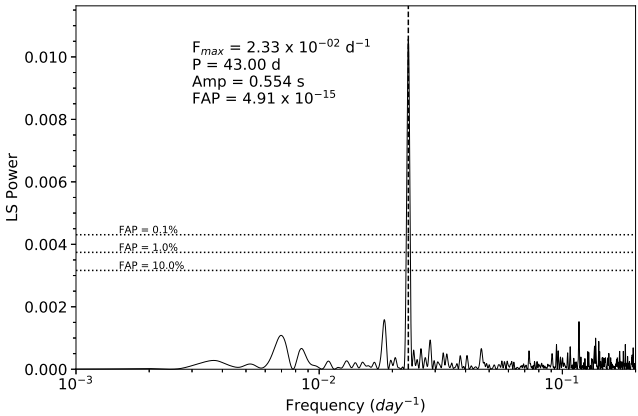}
    \caption{Lomb-Scargle spectrum of the residuals of mid-eclipse timing data from the fitting of two planet solution. The highest peak corresponds to 43 d periodic signal with almost zero False Alarm Probability.}
    \label{fig:LS_spec}
\end{figure}

We then continued and searched for signs of periodicity in the residuals of the two planet solution in order to check if we are under-fitting since the  $\chi_\nu^2$ is still large. For this purpose, we used our own script that makes use of the astropy \citep{astropy2013, astropy2018} module's LS periodogram tool \citep{vanderPlas2018}. We ignored the binary's orbital frequency and the 3.7 d signal by limiting the range of frequency spectrum to a maximum of 0.2 d$^{-1}$.

\begin{figure*}
    \includegraphics[width=460pt]{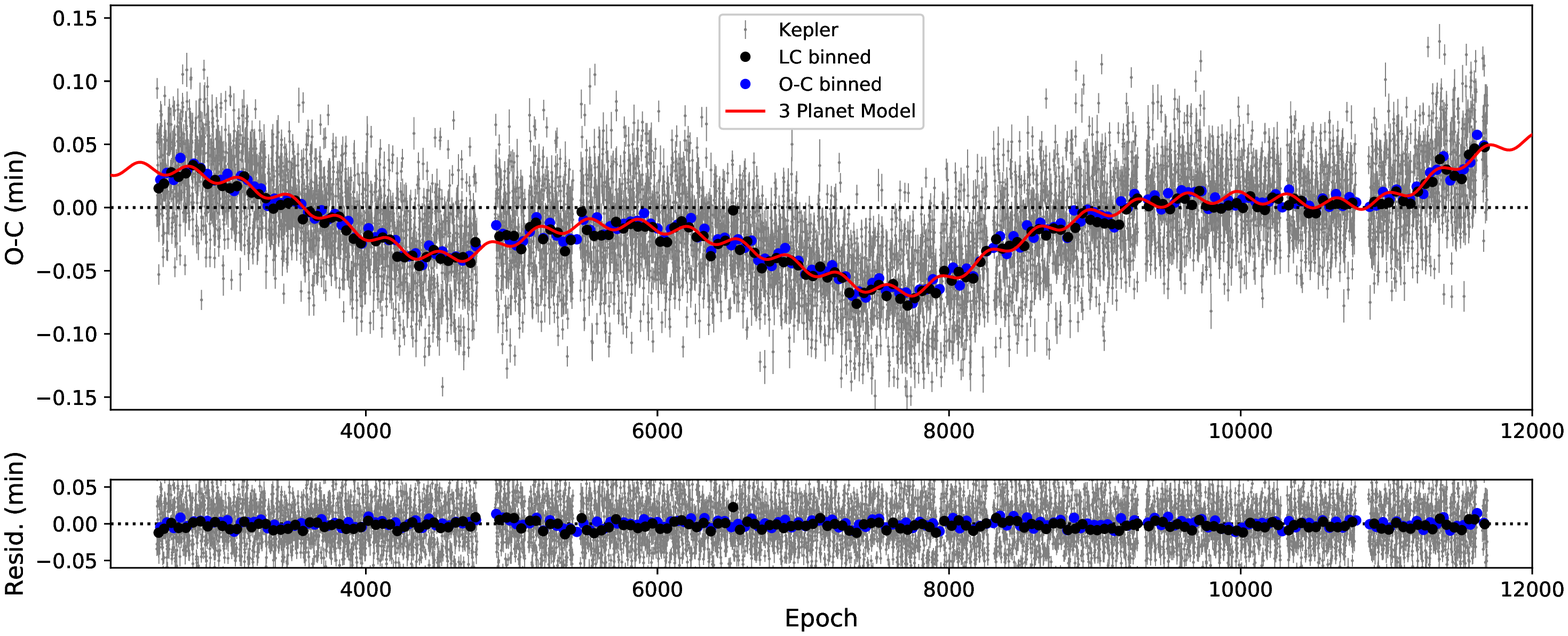}
    \caption{Three planet solution of the mid-eclipse modeling of Kepler-451 in a close-up view to the Kepler data, excluding Quarter 0. Black circles represent the mid-eclipse timings from light curve model fitting to 50 times binned light curves. Blue circles represent the binned O-C data. Note that both binned data follows the trend of the 43 d modulation.}
    \label{fig:oc_model_kepler}
\end{figure*}

We first analysed the primary mid-eclipse timings of Kepler data. The peak frequency in the LS power spectrum (Fig. \ref{fig:LS_spec}) corresponds to a periodicity at 43.0 d. This peak has a False Alarm Probability (FAP) of $4.91 \times 10^{-15}$, and an amplitude of 0.554 s. The same peak exists in the Kepler secondary timings with almost the same amplitude, however, FAP of the peak for the secondaries is unity. Next, we searched for the same frequency in the TESS subset. The total time-span of sector 14 and 15 is 54 d. Sector 40 is separated from the other two sectors by around 1.8 years. So, we searched for potential frequencies only in sectors 14 and 15. The frequency with the highest power corresponds to $\sim$ 39 d then, which is close to what we have found from the Kepler subset. However, due to relatively larger scatter and timing uncertainties, the FAP value is almost unity. The amplitude, on the other hand is around 20 times larger, again, compared to the Kepler subset. We think that the highest peak in TESS spectrum is likely to be not related to 43 d signal that we found in frequency spectrum of the Kepler primary minima.

We investigated if there is an improvement in the ETV fit when we add the 43-day frequency to our model as an additional LiTE. We set the initial values of the period to 43 d and the semi-amplitude (A$_{\rm LiTE,3}$) to 0.5 s. Since the amplitude is too low to model eccentricity, considering timing uncertainties, we assumed a circular orbit for this putative third planet. We also kept every other parameter related to the secular change and the previous two planet solution to their best-fit values. The only free parameters of this model were the period, the amplitude and the mean anomaly for the third LiTE, the best-fit values of which, corresponds to a minimum mass of 1.76 M$\rm _{jup}$ orbiting the binary at a minimum separation of 0.2 au. The $\chi_\nu^2$ value reduced from 12.625 to 12.502 with the addition of the third LiTE

\begin{figure}
    \includegraphics[width=\columnwidth]{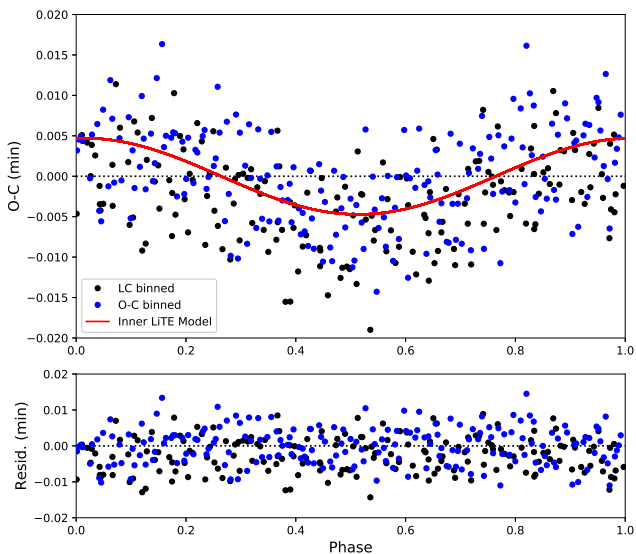}
    \caption{Phased O-C signal corresponding to the third planet model with the period of 43 d and its residuals. Only the binned dataset plotted for clarity.}
    \label{fig:oc_model_phased}
\end{figure}

We phase folded the Kepler light curves to 5, 10, 20 and 50 cycles and recalculated the mid-eclipse times with light curve fitting as we did for the original timings. We confirm that the 43 d signal is present on all of the binned datasets. We plotted only the 50 cycle binned mid-eclipse timings in Fig \ref{fig:oc_model_kepler} as well as binned original Kepler timings with the same number of data points to make the third planet signal for better illustration. Finally, we plotted the binned datasets phased to
the third planet period in Fig \ref{fig:oc_model_phased}.

The final search for a periodicity in the residuals of this three-planet model; however, resulted in no-detection with a significant FAP value. Therefore, our final ETV solution consists of three similar-mass Jovian planets orbiting the binary, with at least two of them on eccentric orbits (Table-\ref{tab:etv_results}). We investigate if such a system will be dynamically stable in the next section.

\begin{table*}
    \centering
    \caption{Results from the eclipse timing variation analysis of Kepler-451. The values in parentheses are uncertainties on the last digits. *Eccentricity of the inner planet is assumed to be zero. **The time and the argument of the periastron of the inner planet are correlated and arbitrary. $n$ denotes the planet number from inside out while the planet names are given in parentheses.}
    \begin{tabular}{c|c|c|c|c}
        \textbf{Parameter} &\textbf{Unit}& \textbf{n=1 (d)} & \textbf{n=2 (b)} & \textbf{n=3 (c)} \\ \hline
        T$_{\rm 0,bin}$ &$\rm BJD_{TDB}$& \multicolumn{3}{c}{2454953.643182(4)}\\
        P$_{\rm0,bin}$ &d&\multicolumn{3}{c}{0.125765285(1)}\\
        $\beta$ & d cycle$^{-2}$&\multicolumn{3}{c}{$-2.9(4) \times 10^{-13}$}\\
        T$_{\rm0,n}$&$\rm BJD_{TDB}$&2454960**&2455131(11)&2456211(38)\\
        P$_n$&$d$&43.0(1)&406(4)&1460(90)\\
        e$_n$&$-$&0*&0.33(5)&0.29(7)\\
        A$_{\rm LiTE,n}$&s&0.28(3)&1.31(3)&2.6(2)\\
        $\omega_n$&$^\circ$&85**&302(8)&7(14)\\
        f(m$_n$)&M$\rm _{jup}$&$1.4(4) \times 10^{-5}$ &$1.6(1) \times 10^{-5}$&$1.1(3) \times 10^{-5}$\\
        m$_n \sin i$&M$\rm _{jup}$&1.76(18)&1.86(5)&1.61(14)\\
        a$_n \sin i$&au&0.20(3)&0.90(4)&2.1(2)\\
    \hline
    \end{tabular}
    \label{tab:etv_results}
\end{table*}

\section{Dynamical Stability of three planet model}\label{sec:stability}

We made an orbital simulation that corresponds to the best-fit parameters to check if the system keeps the initial orbital configuration for a long duration. We assumed all the objects in the system to be co-planar. We also assumed the binary as a single object with a total mass of 0.6 M$_\odot$, which is a very good assumption considering the significantly shorter orbital period compared to that of even the shortest orbital-period planet. We used the Rebound code \citep{rebound} with WHFAST integrator \citep{whfast} for the simulation. Total time of the simulation was set as 10 Myr with a time step of 0.5 d. In this test the orbital coordinates of the objects chaotically oscillate, therefore the system is unstable.

In order to investigate if a stable solution exists in a parameter space close to the best-fit solution, we performed a Frequency Map Analysis (FMA) \citep{laskar_1990Icar...88..266L, laskar_1993PhyD...67..257L}, similarly to our earlier work in \cite{esmer_hwvir}. The timing data covers fewer orbital cycles of the outermost object, hence the uncertainty on its parameters are larger compared to the others. Therefore, we varied the outermost object's semi-major axis ($a$) and eccentricity ($e$) around the best-fit values, while keeping the parameters of the remaining objects to their best-fit values.

We set the total integration time to $10^6$ d, which is around $\sim 680$ orbital revolution of the outermost object, and we set the time step as 0.5 d. The semi-major axis was changed in the range of $1.6 - 2.6$ au, while the eccentricity was changed between 0 - 0.5. For each integration, we calculated the change of the frequency of the mean motion in two halves of the total integration time, and calculated a normalized stability index, D. In the case of stability the value of D has to satisfy the condition, $\rm \log D < -6$ \citep{correia_2005A&A...440..751C}. For the calculation of the frequencies of the mean motion, we used the TRIP code \citep{trip_Gastineau:2011:TCA:1940475.1940518}.

\begin{figure}
    \includegraphics[width=\columnwidth]{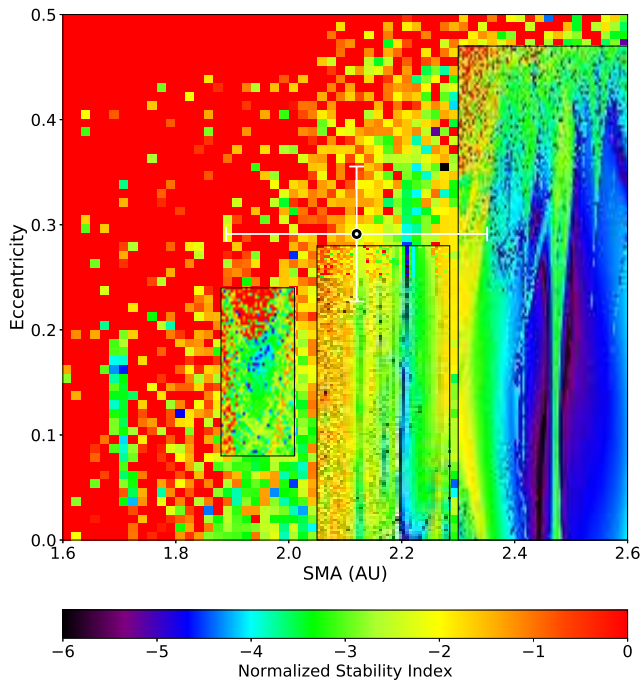}
    \caption{Stability map of the three planet solution of Kepler-451. The central black circle corresponds to the best-fit ETV solution, while the error bars corresponds to 1$\sigma$ uncertainty ranges. Color bar indicates the logarithm of normalized stability index, and the stability occurs when its value is < -6 (black regions). Three rectangles are potentially stable regions. See text for more details.}
    \label{fig:fma_map}
\end{figure}

We initially searched the \emph{a-e} ranges mentioned above in 60 $\times$ 60 resolution. Then, we identified potentially stable sub-ranges and performed the same search for them to reduce the computational time needed for integrations. The three black rectangles shown in Fig. \ref{fig:fma_map} correspond to the potentially stable regions. From left to right, the first region around a = 1.95 au and e = 0.18 has very narrow and scattered  \emph{a-e} coordinates that have normalized stability indices which are closer to the stability condition, none of which actually satisfies it. We think that outermost object is likely to be unstable because it is found to be closer than the best-fit orbital separation to the remaining objects. Therefore, this first candidate region does not represent a stable solution.

The second region around a = 2.2 au and e = 0.15, corresponds to an orbital separation for the outermost object closest to the best-fit solution. This darker and narrow region has very few \emph{a-e} coordinates that actually satisfies the stability condition.

The third region that corresponds to a = 2.5 au has a wide range of stable configurations. The allowed eccentricity values are in a range of 0 - 0.4, for which the upper limit even exceeds the best-fit value for this parameter. The separations of the stable region, on the other hand, are larger around almost 0.3 au (1.5 $\sigma$) from the best-fit value.

We would like to remind that, our eclipse timing solution does not take into account the mutual interactions. One should also keep in mind that the fixed parameters of the outermost object as well as the parameters of the other objects have uncertainties that we did not take into account for the sake of computation time. Our assumption of co-planar orbits is likely to be a good approximation, but there may be slight differences between the mutual orbital inclinations of the objects. Despite all of these, the dynamical stability test shows that stable orbital configurations close to our three planet best-fit eclipse timing solution exist.

\section{Conclusions}\label{sec:conclusions}

Planets around evolved stellar systems are interesting because they may have formed in the accretion disks made up of the ejected material from the stars at some stage of their evolution. Kepler-451 is one such system, with an evolved sdB primary and a dM secondary. We detect two more additional periodic signals in the ETV that we interpreted as planets in the system in addition to the confirmation of the Kepler-451\,b \citep{Baran2015} based on our analysis of the ETVs of the central binary. These two planets have similar minimum masses (m$\rm _c \sin i$ = 1.61 $\pm$ 0.14  M$_{\rm jup}$ and m$\rm_d \sin i$ = 1.76 $\pm$ 0.18 M$_{\rm jup}$) to that of the already found planet-b (m$\rm _b \sin i$ = 1.86 $\pm$ 0.05 M$_{\rm jup}$). Since the sdB primary of Kepler-451 system must have ejected a significant fraction of its mass, these three planetary-mass bodies might be second-generation planets formed in this material. 

The projected orbital separations are a$\rm _d \sin i$ = 0.20 $\pm$ 0.03, a$\rm _b \sin i$ = 0.90 $\pm$ 0.04, and $a\rm _c \sin i = 2.10 \pm 0.02$ astronomical units, respectively. These three similar-mass, Jovian planets of Kepler-451 should be stable on their orbits even they were eccentric, if the outermost planet's orbital separation is $\sim$ 0.3 au larger than our best-fit Keplerian solution, which requires the outermost planet's orbital period to be at least $\sim$ 1800 d. Nevertheless, slight differences in their mutual orbital inclinations and/or eccentricities may lead to other stable configurations too.

The total radial velocity contribution of the three planets then should have a semi-amplitude of $\sim$ 250 m/s depending on their instantaneous mean anomaly, which can be traced with radial velocity observations having achievable precision but on long timescales for verification with an independent method. On the other hand, the orbital inclination of the binary is 70$^\circ$. It is very likely for the circumbinary planets to have similar orbital inclinations. Therefore, they should not show any transits at all, and any information of their radii will be absent until another method is suggested. However, if they have formed from the ejected outer layers of the sdB primary, they should be expected to have very low densities. Therefore, finding planetary mass bodies in transit of their sdB hosts would be very interesting.

The light curve modeling of Kepler-451 showed that the reflection effect on such sdB + dM systems is not trivial to handle. The Kepler light curve and their correlated residuals are not symmetrical when we fold the light curve with respect to the second eclipse, implying that the brightness distribution due to reflection is not homogeneous on the secondaries reflected hemisphere. This can be due to poor-handling of heat re-distribution over the stellar surfaces by the models. However, the effect is small and should have no influence on the determination of the mid-eclipse times.

Calculations of the mid-eclipse timings from the light curve models provide much more reliable data sets for an ETV analysis, as is the usual practice in TTV studies based on model-dependent transit timings. Model-based timings are especially handy in the presence of phase-dependent brightness variations such as spot or pulsation modulations, which pose major problems in traditional methods such as KW due to the light curve asymmetries they cause. 

We would like to mention that the mid-eclipse timings calculated by \cite{Krzesinski2020} from the Kepler light curves are significantly different from the values we derived. We calculated mid-eclipse timings with KW and light curve model fitting from Kepler PDC-SAP light curves and found similar results that only differ in their scatter. However, our mid-eclipse timings do agree with that \cite{Baran2015} derived from the same Kepler light curves, the analysis of which ended up in sinusoidals with very similar parameters that we also attributed to the first circumbinary planet (Kepler-451 b) in the system suggested by them.

Together with the suggested two additional planets in Kepler-451 system, the number of CB planets discovered with the timing technique increases up to 22 when the NASA Exoplanet Archive and exoplanet.eu databases are considered together. Analyzing the eclipse timing variations will continue to play a major role not only in revealing additional circumbinary bodies on larger orbits which allow us to probe a different parameter space, but also study current and past astrophysical phenomena in the evolutions of these systems.

\section*{Acknowledgements}

We gratefully acknowledge the support by The Scientific and Technological Research Council of Turkey (T\"UB\.{I}TAK) with the project 118F042. We thank T\"UB\.{I}TAK for the partial support in using T100 telescope with the project number 19AT100-1471 and 17BT100-1208. This work is also supported by the research fund of Ankara University (BAP) through the project 18A0759001. E.M.E. acknowledges support from T\"UB\.{I}TAK (2214-A, No. 1059B141800521). IST60 telescope and its equipment are supported by the Republic of Turkey Ministry of Development (2016K12137) and Istanbul University with the project numbers BAP-3685, FBG-2017-23943. Some/all of the data presented in this paper were obtained from the Multimission Archive at the Space Telescope Science Institute (MAST). STScI is operated by the Association of Universities for Research in Astronomy, Inc., under NASA contract NAS5-26555. Support for MAST for non-HST data is provided by the NASA Office of Space Science via grant NAG5-7584 and by other grants and contracts. This research has made use of the NASA Exoplanet Archive, which is operated by the California Institute of Technology, under contract with the National Aeronautics and Space Administration under the Exoplanet Exploration Program. This research has made use of data obtained from or tools provided by the portal exoplanet.eu of The Extrasolar Planets Encyclopaedia. We thank the developers of the {\sc phoebe} code, especially Dr. Kyle Conroy from Villanova University for his help in using the code.

\section*{Data Availability}
Eclipse timing measurements and the light curves appearing for the first time in this work are presented as online material in the machine readable format and can be found in the CDS linked to this study. Although we provide the light curves within the online data; we can also provide raw data sets of our own observations on request from the corresponding author.



\bibliographystyle{mnras}
\bibliography{kepler451} 




\appendix

\section{Mid-eclipse times of Kepler-451}

\begin{table}
    \centering
    \caption{Mid-eclipse times of Kepler-451 from our observations.}
    \begin{tabular}{l|l|c|c|c}
       \textbf{Mid-time} &\textbf{Error}&\textbf{Type} & \textbf{Filter}&\textbf{Tel.} \\
       \textbf{$(BJD_{TDB})$} & \textbf{(d)} & \textbf{p/s} \\\hline
2457917.42751	&	0.00011	&	p	&	R	&	TUG	\\
2457917.49022	&	0.00011	&	s	&	R	&	TUG	\\
2458249.57341	&	0.00016	&	p	&	R	&	TUG	\\
2458319.49899	&	0.00008	&	p	&	R	&	TUG	\\
2458635.48425	&	0.00017	&	s	&	R	&	TUG	\\
2458644.41297	&	0.00024	&	s	&	R	&	TUG	\\
2458644.538567	&	0.000059	&	s	&	R	&	TUG	\\
2458798.225663	&	0.000031	&	s	&	R	&	TUG	\\
2458798.226909	&	0.000077	&	s	&	V	&	TUG	\\
2458798.288560	&	0.000084	&	p	&	R	&	TUG	\\
2458798.288606	&	0.000058	&	p	&	V	&	TUG	\\
2458798.35074	&	0.00018	&	s	&	R	&	TUG	\\
2459050.320543	&	0.000023	&	p	&	R	&	IUO	\\
2459130.308316	&	0.000095	&	p	&	R	&	TUG	\\
2459328.514133	&	0.000074	&	p	&	R	&	TUG	\\
2459412.33641	&	0.00022	&	s	&	R	&	IUO	\\
2459412.39981	&	0.00023	&	p	&	R	&	IUO	\\
2459415.48072	&	0.00012	&	s	&	B	&	TUG	\\
2459415.543712	&	0.000089	&	p	&	B	&	TUG	\\
2459417.304485	&	0.000052	&	p	&	V	&	TUG	\\
2459417.36636	&	0.00016	&	s	&	V	&	TUG	\\
2459417.493565	&	0.000058	&	s	&	V	&	TUG	\\
2459425.35294	&	0.00012	&	p	&	R	&	IUO	\\
2459432.458834	&	0.000026	&	s	&	V	&	TUG	\\
2459432.522001	&	0.000048	&	p	&	V	&	TUG	\\
2459443.2749952	&	0.0000098	&	s	&	g’	&	AUKR	\\
2459443.3378487	&	0.0000091	&	p	&	g’	&	AUKR	\\
2459443.401074	&	0.000042	&	s	&	g’	&	AUKR	\\
2459443.463801	&	0.000014	&	p	&	g’	&	AUKR	\\
2459443.526746	&	0.000042	&	s	&	g’	&	AUKR	\\
2459444.280604	&	0.000027	&	s	&	r’	&	AUKR	\\
2459444.344009	&	0.000009	&	p	&	r’	&	AUKR	\\
2459444.406795	&	0.000049	&	s	&	r’	&	AUKR	\\
2459444.4697146	&	0.0000088	&	p	&	r’	&	AUKR	\\
2459444.532454	&	0.000042	&	s	&	r’	&	AUKR	\\
2459447.487990	&	0.000006	&	p	&	i	&	AUKR	\\
2459447.550911	&	0.000013	&	s	&	i	&	AUKR	\\
2459448.2428135	&	0.0000099	&	p	&	g’	&	AUKR	\\
2459448.305610	&	0.000036	&	s	&	g’	&	AUKR	\\
2459448.3685288	&	0.0000075	&	p	&	g’	&	AUKR	\\
2459448.4313756	&	0.0000038	&	s	&	g’	&	AUKR	\\
2459449.311441	&	0.000094	&	s	&	r’	&	AUKR	\\
2459449.37440	&	0.00011	&	p	&	r’	&	AUKR	\\
2459449.43780	&	0.00025	&	s	&	r’	&	AUKR	\\
2459449.50034	&	0.00004	&	p	&	r’	&	AUKR	\\
2459466.47855	&	0.00007	&	p	&	g’	&	TUG	\\
2459476.28840	&	0.00002	&	p	&	g’	&	TUG	\\
2459479.243883	&	0.000037	&	s	&	R	&	TUG	\\
2459479.30665	&	0.00015	&	p	&	R	&	TUG	\\
2459485.469292	&	0.000079	&	p	&	Clear	&	TUG	\\
    \hline
    \end{tabular}
    \label{tab:timings}
\end{table}


\bsp	
\label{lastpage}
\end{document}